\documentclass{article}
\usepackage[onecolumn]{emulateapj}

\newcommand{\lap}{\lesssim}
\newcommand{\gap}{\gtrsim}

\newcommand{\msun}{M_\odot}

\newcommand{\beq}{\begin{equation}}
\newcommand{\eeq}{\end{equation}}

\newcommand{\vp}{V_{\rm plunge}}

\def\mf{m}
\def\nf{n}
\def\ff{f_f}
\def\sf{\sigma_f}

\def\vf{u}

\def\vp{v_{\parallel}}
\def\vt{v_{\perp}}

\def\sstar{\sigma_\star}
\def\df{\langle\Delta\vp\rangle}
\def\dvp{\langle\Delta\vp^2\rangle}
\def\dvt{\langle\Delta\vt^2\rangle}

\begin{document}

\title{A Note on Gravitational Brownian Motion}

\author{David Merritt}
\affil{Department of Physics, 
Rochester Institute of Technology,
Rochester, NY 14623 drmsps@ad.rit.edu}

\begin{abstract}
Chandrasekhar's theory of stellar encounters predicts a 
dependence of the Brownian motion of a massive particle 
on the velocity distribution of the perturbing stars.
One consequence is that the expectation value of the massive
object's kinetic energy can be different from that of the 
perturbers.
This effect is shown to be modest however, and substantially smaller 
than claimed in a recent study based on an approximate 
treatment of the encounter equations.

\end{abstract}


\section{Introduction}

A massive object at the center of a stellar system undergoes
a random walk in momentum space as its motion is perturbed
by gravitational encounters with nearby stars.
After the decay of transients, the steady-state velocity distribution
$f(v)$ of the massive object is a balance between the accelerating
forces of random perturbations and the decelerating
force of dynamical friction.
Chandrasekhar's (1942,1943) theory of gravitational encounters
provides expressions for both forces as functions of the
mass and velocity of the test particle, and the
masses and velocity distribution of the field stars.
In the case that the latter have a Maxwellian velocity
distribution, the steady-state $f(v)$ describing the motion
of the massive object is also a Maxwellian,
with rms velocity given by the equipartition relation
\beq
v_{rms}^2 = {m\over M}u_{rms}^2,
\label{eq:equipart}
\eeq
where $M$ and $v$ refer to the massive object and
$m$ and $u$ to the perturbers.

One difference between Brownian motion in gravitating and
non-gravitating systems is that the velocity distribution
$\ff(u)$ of the perturbers in a galaxy or star cluster can be
substantially non-Maxwellian.
The steady-state velocity distribution of the massive object
is still predicted to be Maxwellian (Merritt 2001),
but there is no reason for $v_{rms}$ to satisfy equation 
(\ref{eq:equipart}).
In this note, the steady-state $f(v)$ 
is derived for a massive particle that is subject
to gravitational perturbations from field stars
with an arbitrary distribution of velocities.
Departures from equipartition are found to be
small, even when the background is strongly non-Maxwellian.
The stronger dependence of $v_{rms}$ on $\ff(u)$
claimed in a recent study (Chatterjee, Hernquist \& Loeb 2002, 
hereafter CHL) is shown to result from approximations
made in the application of Chandrasekhar's theory.

\section{Steady-State Velocity Distributions}

The steady-state velocity distribution $f(v)$ of a massive
object (``black hole'')  that interacts via gravitational scattering with stars
is given by the time-independent solution of the Fokker-Planck
equation.
Following the treatment in Merritt (2001) (hereafter Paper I), 
we can write
\beq
0=f\left[\df + {1\over 2v}\left(\dvt - 2\dvp\right)\right] - {1\over 2}{\partial\over\partial v}\left(f\dvp\right)
\label{eq:fp}
\eeq
where $\df$, $\dvp$ and $\dvt$ are the standard diffusion coefficients
describing the motion of the black hole.
In a steady state, we expect that $v$ will be of order
$\sqrt{m/M}\ll 1$ times the typical stellar velocity.
Expanding the diffusion coefficients about $v=0$,
\begin{mathletters}
\begin{eqnarray}
\df &=& -Av+ Bv^3\ldots , \label{expand1} \\
\dvp &=& C + Dv^2\ldots , \label{expand2} \\
\dvt &=& 2(C + Fv^2)\ldots\label{expand3},
\label{123}
\end{eqnarray}
\end{mathletters}
substituting into equation (\ref{eq:fp}), and retaining the lowest
order terms in $v$ gives 
\begin{mathletters}
\begin{eqnarray}
0&\approx& (A+2D-F)f + {C\over 2}{\partial f\over\partial v}{1\over v} \\
&\approx& Af + {C\over 2}{\partial f\over\partial v}{1\over v} 
\end{eqnarray}
\end{mathletters}
since $A$ is of order $M/m$ times $D$ and $F$.
This has solution
\begin{mathletters}
\begin{eqnarray}
f(v) & = & f_0 e^{-v^2/2\sigma^2},\label{soln1} \\
\sigma^2 & = & {C\over 2A},\label{soln2}
\end{eqnarray}
\end{mathletters}
i.e. the black hole's velocity follows a Maxwellian distribution
with 1D velocity dispersion $\sigma$.

If the background stellar
distribution is assumed to be infinite and homogeneous,
the diffusion coefficients are (Paper I)
\begin{mathletters}
\begin{eqnarray}
\df &=& -16\pi^2 G^2 M\mf \nf \int_0^{\infty}d\vf \left({\vf\over v}\right)^2 \ff(\vf) H_1(v,\vf,p_{max}),
\label{eq:df} \\
\dvp & = & {32\over 3}\pi^2 G^2\mf^2\nf v \int_0^{\infty}d\vf \left({\vf\over v}\right)^2 \ff(\vf) H_2(v,\vf,p_{max}), 
\label{eq:dvp}
\end{eqnarray}
\end{mathletters}
\noindent
where $\ff(u)$ is the velocity distribution of the field stars,
normalized to unit number; $n$ is the field star density;
$p_{max}$ is the maximum impact parameter at which incoming stars
are assumed to perturb the black hole; and
\begin{mathletters}
\begin{eqnarray}
H_1(v,\vf,p_{max}) &=& {1\over 8\vf}\int_{|v-\vf|}^{v+\vf} dV\left(1 + {v^2-\vf^2\over V^2}\right) \ln\left(1 + {p^2_{max}V^4\over G^2 M^2}\right),
\label{eq:h1}\\
H_2(v,\vf,p_{max}) & = &{3\over 8\vf}\int_{|v-\vf|}^{v+\vf} dV \Biggl\{\left[1-{V^2\over 4v^2}\left(1 + {v^2-\vf^2\over V^2}\right)^2\right]\ln\left(1 + {p^2_{max}V^4\over G^2 M^2}\right) \label{eq:h2} \\ 
& + & \left[{3\over 4}{V^2\over v^2}\left(1+{v^2-\vf^2\over V^2}\right)-1\right]{p^2_{max}V^4/G^2M^2\over 1+p^2_{max}V^4/G^2M^2}\Biggr\} \nonumber
\end{eqnarray}
\end{mathletters}
Equation (\ref{eq:h2}) includes the ``non-dominant'' 
terms which are of order unity when $p_{max}$ is large.
The integration variable $V$ is the relative velocity of star and black
hole when the star is at infinity.

In the standard approximation (e.g. Rosenbluth, MacDonald \& Judd 1957;
Spitzer 1987), 
the non-dominant terms are neglected, 
and the logarithmic terms are taken outside the integrals,
since they are slowly varying with respect to $V$.
One writes
\beq
\ln\left(1+{p^2_{max}V^4\over G^2M^2}\right) \approx 2\ln\Lambda,
\eeq
a constant, and 
\begin{mathletters}
\begin{eqnarray}
\df &=& -16\pi^2 G^2 M\mf\nf\ln\Lambda \int_0^v d\vf \left({\vf\over v}\right)^2 \ff(\vf) ,\label{eq:df1}\\
\dvp & = & {32\over 3}\pi^2G^2\mf^2\nf\ln\Lambda v\left[ 
\int_0^v d\vf \left({\vf\over v}\right)^4 \ff(\vf) +  
\int_v^\infty d\vf \left({\vf\over v}\right) \ff(\vf) \right],
\end{eqnarray}
\end{mathletters}
with leading terms
\begin{mathletters}
\begin{eqnarray}
A_\Lambda &=& {16\over 3} \pi^2G^2 M\mf\nf\ln\Lambda \ff(0), \\
C_\Lambda &=&  {32\over 3}\pi^2G^2\mf^2\nf\ln\Lambda \int_0^\infty d\vf \vf \ff(\vf).
\end{eqnarray}
\end{mathletters}
The predicted 1D velocity dispersion of the black hole is then (CHL)
\beq
\sigma_\Lambda^2 = {C_\Lambda\over 2A_\Lambda} = \left({m\over M}\right) 
\ff(0)^{-1}\int_0^{\infty}d\vf\vf \ff(\vf).
\label{eq:CHL}
\eeq

However when the velocity of the massive particle is very low,
the logarithms in equations (\ref{eq:h1}-b) will be close
to zero for {\it all} stars with $\vf<v$, i.e. all stars that 
enter into the dynamical friction integral (\ref{eq:df1}).
It is unclear in this case whether removing the logarithm 
from the integrands is a reasonable approximation
(Chandrasekhar 1943; White 1949).

Returning to the exact expressions (\ref{eq:df}-\ref{eq:h2}) 
and expanding to lowest order in $v$, we find after some algebra:
\begin{mathletters}
\begin{eqnarray}
A &=& {32\over 3}\pi^2 G^2 M\mf\nf \int_0^{\infty} {d\vf\over\vf} 
\ff(\vf) {p_{max}^2\vf^4/G^2M^2\over 1+p_{max}^2\vf^4/G^2M^2}, \\
C &=&  {16\over 3}\pi^2G^2\mf^2\nf \int_0^\infty d\vf\vf\ff(\vf)
\log\left(1+{p_{max}^2\vf^4\over G^2M^2}\right).
\end{eqnarray}
\end{mathletters}
The non-dominant terms may be shown not to contribute to this order.
Note that both diffusion coefficients now depend on the field-star
velocity distribution at {\it all} values of $u$, and in fact
the field stars that produce the dynamical friction force can be shown 
to have roughly the same velocity distribution as $\ff(u)$ (Paper I).
In the standard treatment with ``$\log\Lambda$'s removed,''
the dynamical friction force comes entirely from stars with
$u<v$ (as in equation~\ref{eq:df1}), a physically unreasonable
result.

The black hole's velocity dispersion is now
\beq
\sigma^2 = {C\over 2A} = {1\over 4}{m\over M}
{\int_0^\infty d\vf\vf\ff(\vf) \log\left(1+{p_{max}^2\vf^4\over G^2M^2}\right) 
\over 
\int_0^{\infty} d\vf \vf^{-1} 
\ff(\vf) {p_{max}^2\vf^4/G^2M^2\over 1+p_{max}^2\vf^4/G^2M^2}}.
\label{eq:exact1}
\eeq
If $\ff(\vf)$ is differentiable at low and high $\vf$,
the denominator can be integrated by parts yielding
\beq
\sigma^2 = {m\over M}
{\int_0^\infty d\vf\vf\ff(\vf) \log\left(1+{p_{max}^2\vf^4\over G^2M^2}\right) 
\over 
\int_0^{\infty} d\vf \left(-d\ff/ d\vf\right) 
\ff(\vf)  \log\left(1+{p_{max}^2\vf^4\over G^2M^2}\right)}.
\label{eq:exact2}
\eeq
This reduces to the approximate expression, equation (\ref{eq:CHL}), 
if the logarithmic
terms are ``taken out of the integrals.''

In the case of a Maxwellian field star velocity distribution,
$\ff(\vf) = f_M(\vf) = f_0 e^{-\vf^2/2\sf^2}$, 
the exact coefficients are
\begin{mathletters}
\begin{eqnarray}
A_M& = & {4\sqrt{2\pi}\over 3} {G^2M m\nf\over\sigma_f^3} G(R_M), \ \ \ \ \ \ 
C_M = {8\sqrt{2\pi}\over 3} {G^2\mf^2\nf\over\sf} G(R_M),\\
G(R_M) & \equiv & {1\over 2} \int_0^{\infty} dz\ e^{-z} \ln\left(1+4 R_M^2z^2\right), \ \ \ \ \ \ R_M \equiv {p_{max}\sf^2\over GM}.\label{eq:gofr}
\end{eqnarray}
\end{mathletters}
\noindent
and
\beq
\sigma^2 = {C_M\over 2A_M}=\left({m\over M}\right)\sf^2,
\eeq
independent of $p_{max}$ (Paper I).
Remarkably, this is also the value of $\sigma^2$ 
given by the approximate expression
 (\ref{eq:CHL}) when $\ff(u)=f_M(u)$.

For all other -- non-Maxwellian -- $\ff$'s, however, 
equations (\ref{eq:exact1}) 
and (\ref{eq:CHL}) yield different results for
$\sigma^2$, and each is different from the Maxwellian value.
Furthermore $\sigma^2$ as predicted by the exact
expression (\ref{eq:exact1}) is in general a function 
of the maximum impact parameter, since
$C$ and $A$ depend differently on $p_{max}$.

Without loss of generality, we can express the dependence
of $\sigma^2$ on $p_{max}$ via a dimensionless
variable $R$, where $R\equiv p_{max}\tilde{\sigma}^2/GM$ and
$\tilde{\sigma}$ is a 1D velocity dispersion that characterizes
the field star velocity distribution.
Equating $G(R)$ in equation~(\ref{eq:gofr}) 
with $\ln\Lambda\equiv\ln (p_{max}/p_{min})$ reveals that 
$p_{min}\approx GM/\sqrt{2}\sf$ (Paper I), 
hence $R\approx p_{max}/p_{min}$; note that $p_{min}$
is also approximately equal to the ``radius of influence'' 
of the black hole.
When $p_{max}$ is large, i.e. $R\gg 1$,
the logarithmic terms
in equation (\ref{eq:exact2}) are nearly constant 
and $\sigma^2$ reduces to the value (\ref{eq:CHL}) 
derived by ``taking out $\ln\Lambda$.''
But for a supermassive black hole in a galactic nucleus, 
$p_{max}$ is of order the core radius $r_c$ of the 
stellar distribution (Paper I),
and observed core radii are of order a few times 
$GM/\sf^2$ at most (e.g. Poon \& Merritt 2001); 
hence $p_{max}\gap p_{min}$
and $R$ is of order unity.
Thus we expect the exact and approximate expressions
for $\sigma^2$ to be substantially different in many
astrophysically interesting situations.

\section{Examples}

The equations derived above can be applied to various field-star
distributions and the results for $\sigma$ compared
with the equipartition value.
Following CHL, we quantify
the departure of $\sigma$ from equipartition via
the parameter $\eta$, defined as
\beq
\eta \equiv {M\over m}{\sigma^2\over \sstar^2}
\eeq
with $\sstar$ the stellar velocity dispersion at the location 
of the black hole;
$\eta$ is unity in the case of equipartition.
Both $\sigma$ and $\sstar$ are computed under
the assumption that the black hole sits at the
center of the stellar system.
We emphasize that the effect of the
black hole on the stellar background is ignored.

King's (1966) family of models
\beq
f_K(E)=\cases{(2\pi\sigma_K^2)^{-3/2}\left(e^{E/\sigma_K^2}-1\right),&if $E\ge 0$;\cr
0,&otherwise\cr}
\eeq
have nearly Maxwellian central velocity distributions.
Here $E$ is the binding energy, $E=-\vf^2/2+\psi(r)$, 
and $\psi(r)=-\Phi(r)$ with $\Phi(r)$
the gravitational potential due to the stars, 
equal to zero at the model's edge.
The parameter $W\equiv\psi(0)/\sigma_K^2$ is a dimensionless
measure of the central potential.
For large $W$, 
the King model parameter 
$\sigma_K$ is nearly equal to the true central
velocity dispersion $\sstar$, or
\beq
\sstar^2 = {2\over 3} {\int_0^W dE\left(W-E\right)^{3/2}f_K(E)\over
\int_0^W dE\sqrt{W-E}f_K(E)}.
\eeq

A natural definition for $R$ is
\beq
R_K\equiv {p_{max}\sigma_K^2\over GM}.
\label{eq:rk}
\eeq
King models having $W\gap 10$ 
fit the observed brightness profiles of some elliptical 
galaxies moderately well.
Setting $M$ to $\sim 0.1\%$ of the total stellar mass
and equating $p_{max}$ to the core radius gives
$R_K$ of order unity when $W\gap 10$.
Figure 1 shows that $\eta$ is extremely close to 
one when $R$ and $W$ have these values.
King models with smaller $W$ do not represent
galaxies well -- their cores are too large -- 
but are good fits to some globular clusters.
The appropriate value of $R_K$ for globular clusters
is unclear however since it is not known whether such systems
harbor massive central objects or what their likely masses are.
If we suppose that the black hole 
has a mass equal to a fraction $F$ of the stellar
mass in the core,
then $R_K\approx F^{-1}$.
An upper limit on $R$ comes from assuming that the
black hole has a minimum mass $\sim 10\msun$ and that the core mass
is $\sim 5\times 10^4\msun$, its value in a massive, large-cored
globular cluster like $\omega$ Centauri (e.g. Merritt, Meylan \& Mayor 1997).
Then $F\approx 2\times 10^{-4}$ and $R_K\approx 5000$.
In such an extreme case, Figure 1 shows that $\eta$ 
can approach its asymptotic limit of $1.75$ (CHL),
but only in highly unphysical models with $W\lap 2$.

Next we consider the polytropic models, some of which
have strongly non-Maxwellian velocity distributions.
The Plummer (1911) polytrope has $\ff=f_5(E)$, where
\beq
f_n(E) = \cases{f_0E^{n-{3\over 2}},&if $E>0$;\cr
0,&otherwise\cr}
\eeq
and the gravitational potential is again defined to be
zero at the model's edge, which in this case lies at infinity.
The density profile is
\beq
\rho(r) = {3M_{gal}a^2\over 4\pi}{1\over \left(r^2+a^2\right)^{5/2}}
\eeq
and the central velocity dispersion is $\sstar^2(0) = GM_{gal}/6a$.
We define
\beq
R_P \equiv {p_{max}\sstar^2(0)\over GM}
\label{eq:rp}
\eeq
which gives
\beq
\eta = {6\over 7}{\int_0^2dx\left(2-x\right)^{7/2}\log\left(1+36R^2x^2\right)\over \int_0^2 dx\left(2-x\right)^{5/2}\log\left(1+36R^2x^2\right)}.
\eeq
As $R\rightarrow\infty$, $\eta\rightarrow 4/3$ (CHL).
Figure 2 shows $\eta$ as a function of $R$.
Values of $R$ appropriate to supermassive black holes in 
galactic nuclei again give $\eta$ close to one and the asymptotic
value is approached only slowly as $R$ is increased.

The third example considered by CHL was the $n=1$ polytrope.
The structure of this model makes it a poor representation
of {\it any} stellar system, since the density profile:
\beq
\rho(r) = {M_{gal}\over 4\pi^2a^2}r^{-1}\sin(r/a),\ \ \ \ r<\pi a
\eeq
is ``all core.''
Furthermore the distribution function is singular
at the model's edge.
For this reason we must use the first of the two
expressions derived above for $\sigma^2$,
equation (\ref{eq:exact1}), since the second assumes
differentiability of $\ff$.
We define $R$ in terms of the central velocity dispersion
as in equation (\ref{eq:rp}),
where $\sigma_\star^2(0)=GM_{gal}/2\pi a$.
The result for $\eta$ is
\beq
\eta = {1\over 16\pi R^2}{\int_0^{1/\pi} dx\left({1\over\pi}-x\right)^{-1/2}\log\left(1+16\pi^2 R^2\right)\over \int_0^{1/\pi} dx x \left({1\over\pi}-x\right)^{-1/2}\left( 1+16\pi^2 R^2\right)^{-1}}.
\eeq
Figure 3 plots this function.
The asymptotic value $\eta=4$ (CHL) is again reached only very gradually
as $R$ increases.

\section{Discussion}

The foregoing should not be taken to imply that
a massive object at the center of a stellar system
should always be in or near a state of equipartition 
with respect to its less-massive perturbers.
Indeed it is not even clear what Chandrasekhar's 
theory would predict in many physically interesting
situations.
Chandrasekhar's theory is local in the sense that the
density and velocity distribution of the field stars
are assumed to everywhere be the same as they are at the location
of the test particle.
The effect of a massive particle on the structure of the
background stellar system is also ignored.
Both of these assumptions make it difficult to
apply the theory to the case of 
Brownian motion of a supermassive
black hole at the center of a galactic nucleus.
The density of stars falls steeply 
away from the black hole in most galaxies; indeed a density falloff
as steep as $\rho\sim r^{-1/2}$ is required for
self-consistency if the stellar velocities are
isotropic near the black hole (as assumed here).
Furthermore the presence of the black hole implies
a steeply-rising stellar velocity dispersion 
within its sphere of influence, $r\lap GM/\sigma_\star^2$,
and the black hole's Brownian motion would be influenced
by these higher velocities.

The likely importance of such effects is suggested by 
a recent numerical study of Brownian motion
in a galaxy with a steep, $\rho\sim r^{-3/2}$
density cusp (Dorband, Hemsendorf \& Merritt 2003).
The stellar velocity dispersion in this model
has a central value of zero in the absence of the
black hole, yet the numerical integrations reveal
a significant Brownian motion of the ``black hole.''
Evidently
the massive particles in these simulations
are responding to stars far from the center,
or to stars whose motions have
themselves been influenced by the presence of the
black hole, or both.

\bigskip

This work was supported by grants AST-0071099 and 
AST-0206031 from the NSF and by grant MAG5-9046
from NASA.

\begin{figure}
\epsscale{0.6}
\plotone{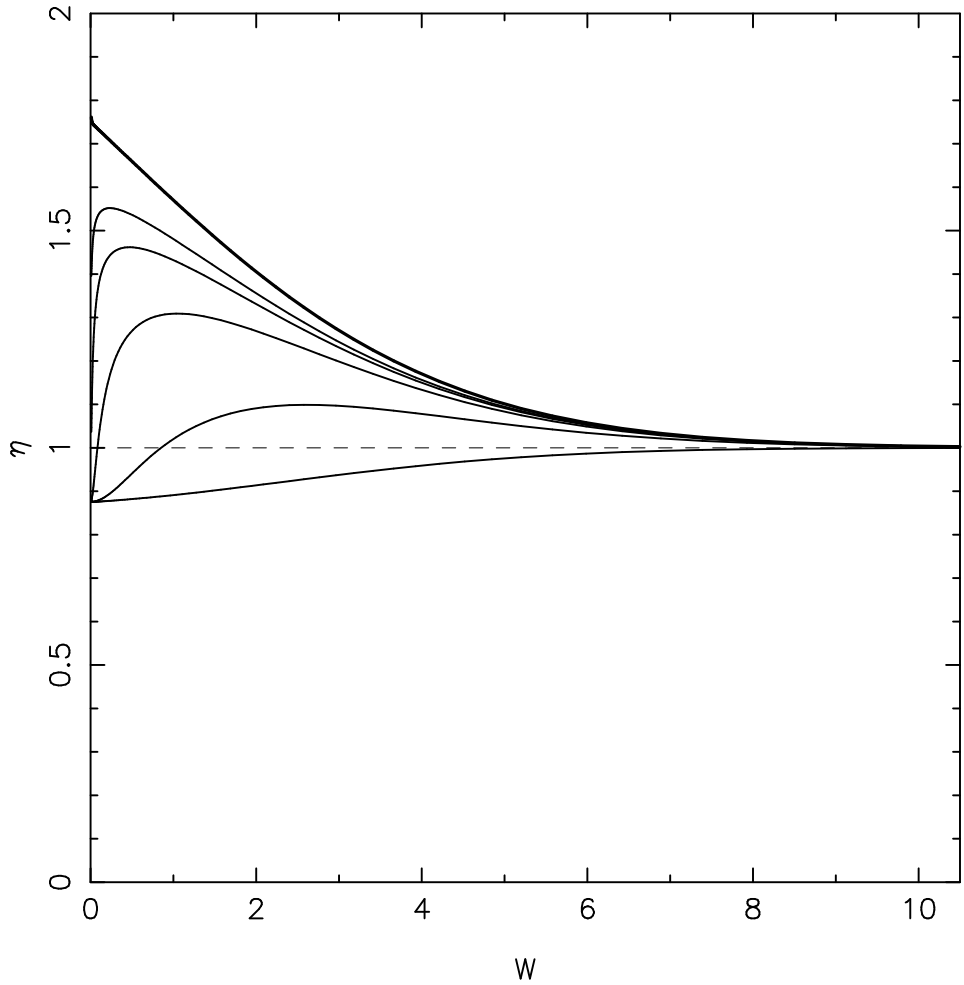}
\caption{\label{fig:king}
Departure from equipartition of a massive object at the
center of a King-model galaxy.
Thin curves are for $R=0.1,1,10,100, 1000$ moving upward,
where $R$ is the dimensionless maximum impact parameter
defined by equation~(\ref{eq:rk}).
Thick curve is the asymptotic form as $R\rightarrow\infty$.
$R$ is expected to be of order unity for a supermassive
black hole at the center of a galaxy.
}
\end{figure}

\begin{figure}
\plotone{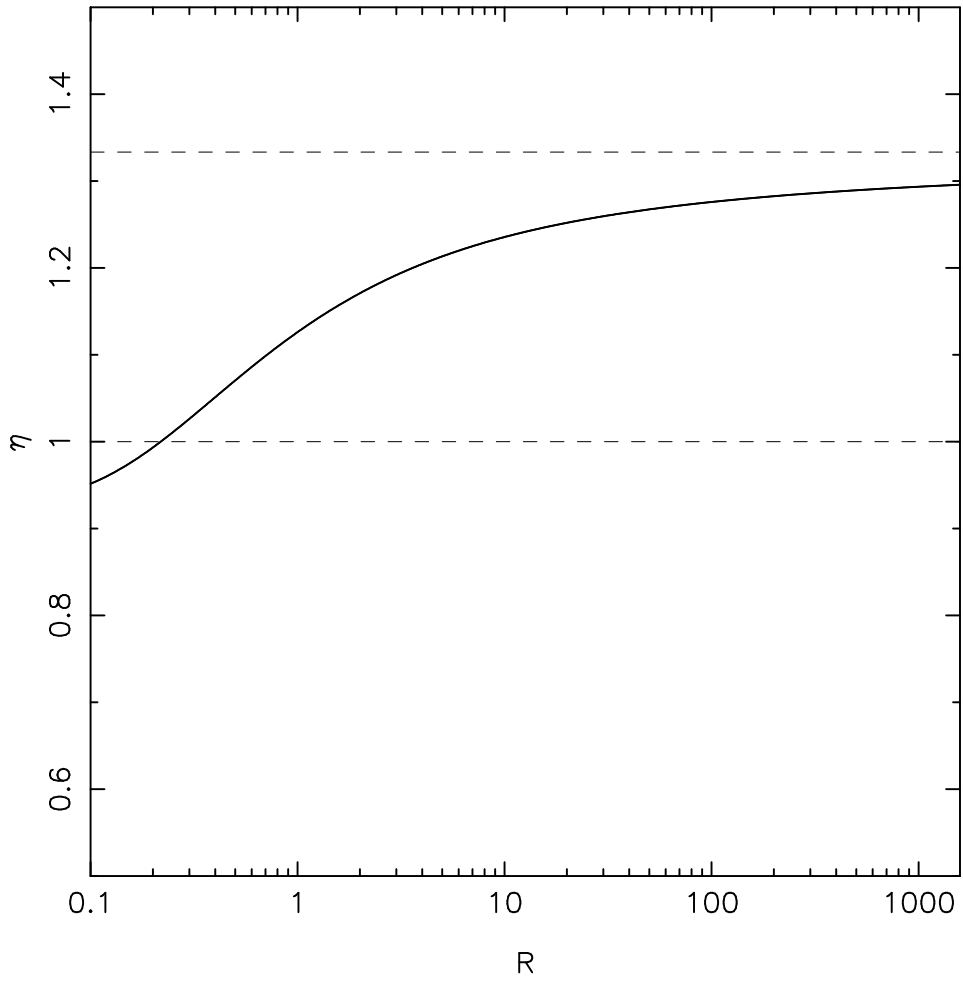}
\caption{\label{fig:plummer}
Values of $\eta$ for a massive object at the center of a
Plummer-model galaxy, as a function of the dimensionless
maximum impact parameter $R$ of equation~(\ref{eq:rp}).
Upper dashed line shows asymptotic value for large $R$.
}
\end{figure}

\begin{figure}
\plotone{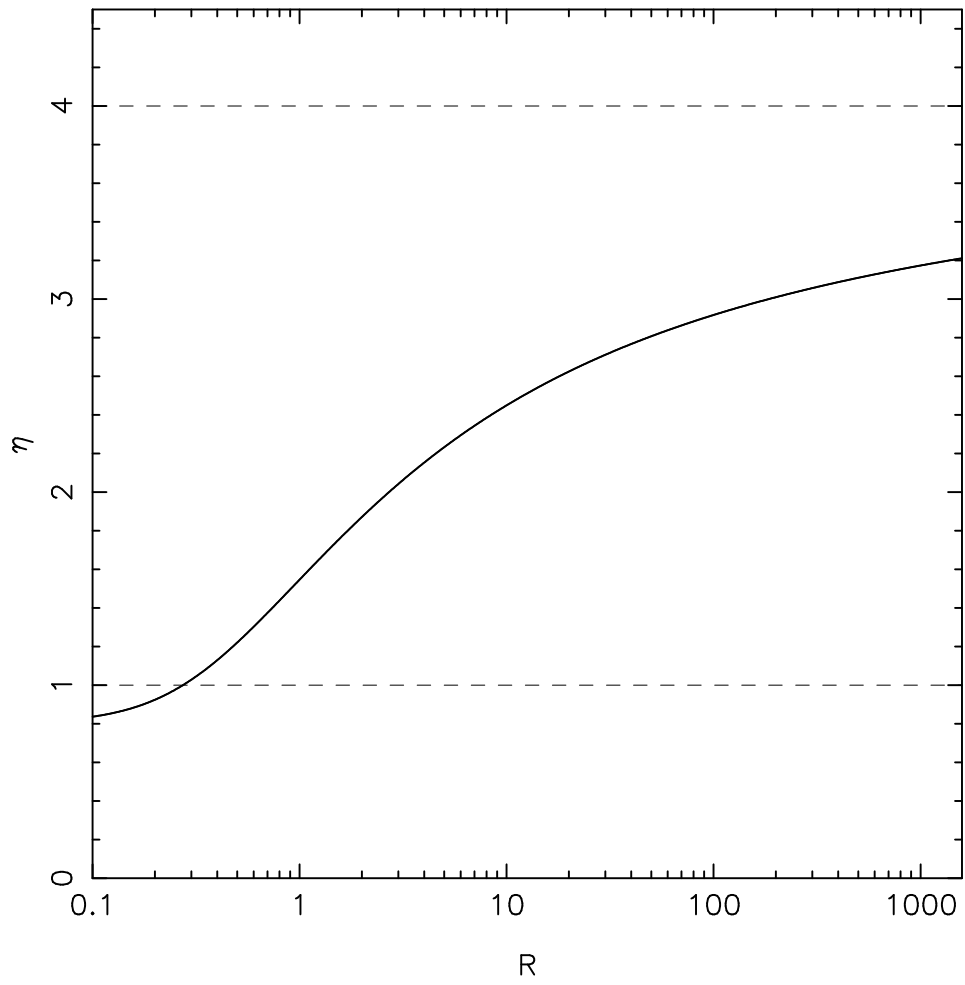}
\caption{\label{fig:poly}
Like Figure 2, for the $n=1$ polytrope.
}
\end{figure}

\end{document}